\documentclass[
aps,
prl,
reprint,
groupedaddress, 
superscriptaddress,
amsmath,
amssymb,
]{revtex4-2}

\usepackage{graphicx}
\usepackage{dcolumn}
\usepackage{bm}
\usepackage{orcidlink}
\usepackage[stretch=10]{microtype}  
\usepackage{hyperref}

\usepackage{float}
\usepackage[export]{adjustbox}
\usepackage{xcolor}

\newcommand{\ja}[1]{\textcolor{black}{#1}}
\begin{document}


\title{\textbf{High-Resolution Laser Spectroscopy on the Hyperfine Structure of $^{255}$Fm (Z=100)} 
}

\author{Mitzi Urquiza-González\orcidlink{0000-0001-6603-4375}}
\email{Contact author: mitzi.urquiza@gu.se}
 \affiliation{University of Gothenburg, Gothenburg, Sweden}
 \affiliation{Division HÜBNER Photonics, Hübner GmbH \& Co KG, Kassel, Germany}

\author{Matou Stemmler}%
\affiliation{Johannes Gutenberg Universität Mainz, Mainz, Germany}
\affiliation{Leibniz University Hannover, Hannover, Germany}

\author{Thomas E. Albrecht}
\affiliation{Florida State University, Tallahassee, FL, USA}

\author{\textcolor{black}{Benjamin Bally\orcidlink{0000-0002-1802-6802}}}
\affiliation{Technische Universit\"at Darmstadt, Darmstadt, Germany}

\author{\textcolor{black}{Michael Bender\orcidlink{0000-0001-8707-3410}}}
\affiliation{Universit{\'e} Claude Bernard Lyon 1, CNRS/INP2P3, IP2I Lyon, UMR 5822, F-69100 Villeurbanne, France}

\author{Sebastian Berndt}
\affiliation{Johannes Gutenberg Universität Mainz, Mainz, Germany}

\author{Michael Block}
\affiliation{Johannes Gutenberg Universität Mainz, Mainz, Germany}
\affiliation{Helmholtz-Institut Mainz, Mainz, Germany}
\affiliation{GSI Helmholtzzentrum für Schwerionenforschung GmbH, Darmstadt, Germany}

\author{Alexandre Brizard}
\affiliation{GSI Helmholtzzentrum für Schwerionenforschung GmbH, Darmstadt, Germany}
\affiliation{GANIL, Caen, France}%

\author{\ja{Joseph S. Andrews}\orcidlink{0009-0005-4535-8342}}
\affiliation{GSI Helmholtzzentrum für Schwerionenforschung GmbH, Darmstadt, Germany}
\affiliation{Helmholtz-Institut Mainz, Mainz, Germany}

\author{\ja{Jacek~Bieroń}\orcidlink{0000-0002-0063-4631}}  
\affiliation{Uniwersytet Jagielloński, 
Kraków, Poland}

\author{Premaditya Chhetri}
\affiliation{Johannes Gutenberg Universität Mainz, Mainz, Germany}
\affiliation{Helmholtz-Institut Mainz, Mainz, Germany}
\affiliation{GSI Helmholtzzentrum für Schwerionenforschung GmbH, Darmstadt, Germany}

\author{Holger Dorrer}
\affiliation{Johannes Gutenberg Universität Mainz, Mainz, Germany}

\author{Christoph E. Düllmann}
\affiliation{Johannes Gutenberg Universität Mainz, Mainz, Germany}
\affiliation{Helmholtz-Institut Mainz, Mainz, Germany}
\affiliation{GSI Helmholtzzentrum für Schwerionenforschung GmbH, Darmstadt, Germany}

\author{Julie G. Ezold}
\affiliation{Oak Ridge National Laboratory, Oak Ridge, TN, USA}

\author{{\color{black}Stephane Goriely}}
\affiliation{Universit\'e Libre de Bruxelles, Brussels, Belgium}

\author{Manuel J. Gutiérrez}
\affiliation{Helmholtz-Institut Mainz, Mainz, Germany}
\affiliation{GSI Helmholtzzentrum für Schwerionenforschung GmbH, Darmstadt, Germany}
\affiliation{Universität Greifswald, Greifswald, Germany}

\author{Dag Hanstorp}
 \affiliation{University of Gothenburg, Gothenburg, Sweden}

\author{Raphael Hasse}
\affiliation{Johannes Gutenberg Universität Mainz, Mainz, Germany}

\author{Reinhard Heinke}
\affiliation{CERN, Geneva, Switzerland}

\author{Korbinian Hens}
 \affiliation{Division HÜBNER Photonics, Hübner GmbH \& Co KG, Kassel, Germany}

\author{{\color{black}Stephane Hilaire}}
\affiliation{CEA, DAM, DIF, 
Arpajon, France}
\affiliation{
Universit\'e Paris-Saclay, 
Bruy\`eres-le-Ch\^atel, France}

\author{Magdalena Kaja}
\affiliation{Johannes Gutenberg Universität Mainz, Mainz, Germany}

\author{Tom Kieck}
\affiliation{Helmholtz-Institut Mainz, Mainz, Germany}
\affiliation{GSI Helmholtzzentrum für Schwerionenforschung GmbH, Darmstadt, Germany}

\author{Nina Kneip}
\affiliation{Johannes Gutenberg Universität Mainz, Mainz, Germany}

\affiliation{Leibniz University Hannover, Hannover, Germany}

\author{Ulli Köster}
\affiliation{Institut Laue-Langevin, Grenoble, France}

\author{ Andrea T. Loria Basto}
\affiliation{Johannes Gutenberg Universität Mainz, Mainz, Germany}
\affiliation{GSI Helmholtzzentrum für Schwerionenforschung GmbH, Darmstadt, Germany}

\author{Christoph Mokry}
\affiliation{Johannes Gutenberg Universität Mainz, Mainz, Germany}
\affiliation{GSI Helmholtzzentrum für Schwerionenforschung GmbH, Darmstadt, Germany}

\author{Danny Münzberg}
\affiliation{Johannes Gutenberg Universität Mainz, Mainz, Germany}
\affiliation{Helmholtz-Institut Mainz, Mainz, Germany}
\affiliation{GSI Helmholtzzentrum für Schwerionenforschung GmbH, Darmstadt, Germany}

\author{Kristian Myhre}
\affiliation{Oak Ridge National Laboratory, Oak Ridge, TN, USA}

\author{Thorben Niemeyer}
\affiliation{Johannes Gutenberg Universität Mainz, Mainz, Germany}

\author{{\color{black}Sophie P\'eru}}
\affiliation{CEA, DAM, DIF, 
Arpajon, France}
\affiliation{
Universit\'e Paris-Saclay, 
Bruy\`eres-le-Ch\^atel, France}

\author{Sebastian Raeder\orcidlink{0000-0003-0505-1440}}
\affiliation{Helmholtz-Institut Mainz, Mainz, Germany}
\affiliation{GSI Helmholtzzentrum für Schwerionenforschung GmbH, Darmstadt, Germany}

\author{Dennis Renisch}
\affiliation{Johannes Gutenberg Universität Mainz, Mainz, Germany}
\affiliation{Helmholtz-Institut Mainz, Mainz, Germany}

\author{Jörg Runke}
\affiliation{Johannes Gutenberg Universität Mainz, Mainz, Germany}
\affiliation{GSI Helmholtzzentrum für Schwerionenforschung GmbH, Darmstadt, Germany}

\author{Samantha K. Schrell}
\affiliation{Oak Ridge National Laboratory, Oak Ridge, TN, USA}

\author{Dominik Studer}
\affiliation{Helmholtz-Institut Mainz, Mainz, Germany}
\affiliation{GSI Helmholtzzentrum für Schwerionenforschung GmbH, Darmstadt, Germany}



\author{Kenneth van Beek}
\affiliation{GSI Helmholtzzentrum für Schwerionenforschung GmbH, Darmstadt, Germany}
\affiliation{Technische Universit\"at Darmstadt, Darmstadt, Germany}

\author{Jessica Warbinek}
\affiliation{Johannes Gutenberg Universität Mainz, Mainz, Germany}
\affiliation{GSI Helmholtzzentrum für Schwerionenforschung GmbH, Darmstadt, Germany}

\author{Klaus Wendt}
\affiliation{Johannes Gutenberg Universität Mainz, Mainz, Germany}

\date{\today}

\begin{abstract}


    We report on high-resolution laser spectroscopy of $^{255}$Fm ($T_{1/2} = 20$h), one of the heaviest nuclides available from reactor breeding. The hyperfine structures in two different atomic ground-state transitions at 398.4\,nm and 398.2\,nm were probed by in-source laser spectroscopy at the RISIKO mass separator in Mainz, using the PI-LIST high-resolution ion source. Experimental results were combined with hyperfine fields from various atomic \textit{ab-initio} calculations, in particular using MultiConfigurational Dirac-Hartree-Fock theory, as implemented in \textsc{grasp18}. In this manner, the nuclear magnetic dipole and electric quadrupole moments were derived to be  $\mu = -0.75(5)\, \mu_\textrm{N}$ and $Q_\textrm{S} = +5.84(13)$\,eb, respectively.
    The magnetic moment indicates occupation of the $\nu$\,7/2[613] Nilsson orbital, while the large quadrupole moment confirms strong, stable prolate deformation consistent with systematics in the heavy actinides. Comparisons with available expectation values from nuclear theory show good agreement, providing a stringent benchmark for the used theoretical models. 
    These results revise earlier data and establish $^{255}$Fm as a reference isotope for future high-resolution studies.

\end{abstract}

\maketitle

\begin{figure}[t]
\includegraphics[width=0.45\textwidth]{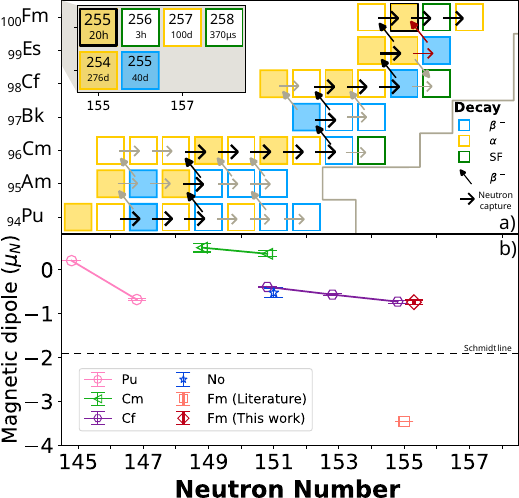}
\caption{\label{fig:chart} a) Section of the nuclear chart showing some of the reactor-accessible nuclides \cite{oakridge2015, breeding-path}, including $^{239}$Pu. Dark arrows show the primary breeding-path, while light arrows show secondary channels. Red arrows indicate the path taken for this work. Fully colored squares highlight isotopes for which nuclear moments are known. 
The main decay modes are color-coded following the key given on the right. An insert shows the end of the production pathway, indicating the nuclides' half-lives. b)~Magnetic dipole moments of even-$Z$ nuclei ranging from plutonium to nobelium \cite{NUBASE2020, Nothhelfer2022, Weber2023, Nobelium_Sebastian}. Close-lying values are offset horizontally for visualization.}
\end{figure}

The radioelement fermium (Fm) with 100 protons in its nucleus is the heaviest element which can be produced in macroscopic amounts. 
Neutron-capture and nuclear beta-decays of lighter actinide isotopes in nuclear reactors enable the production of pg quantities \cite{oakridge2015}, while heavier elements are produced only in atom-at-a-time quantities \cite{smits2024quest}.
Despite its macroscopic availability, atomic information for fermium is limited \cite{Block2021}. 
So far, seven optical transitions have been measured with laser spectroscopy \cite{Sewtz2003, Backe2005}.
This method can unveil fundamental atomic and nuclear structure properties of nuclides via precision studies of their atomic spectra \cite{Yang2023}. 
In this context, isotope shifts, which give access to changes in the nuclear mean-square charge radii along isotopic chains, were recently measured in fermium, spanning from  $^{245}$Fm to $^{257}$Fm \cite{Warbinek2024}. 
Furthermore, investigating the hyperfine structure (HFS) in optical transitions gives access to nuclear moments, which are scarcely known for the heaviest elements, and whose values provide sensitive indicators of underlying shell structure.
This fact is visualized in Fig.~\ref{fig:chart}~a), which summarizes the known nuclear dipole moments for nuclides in the region of the heavy actinides.
A first attempt to probe the hyperfine structure in a fermium isotope was performed on $^{255}$Fm in Ref.~\cite{Backe2005} with limited resolution and the obtained nuclear dipole moment is reported in the relevant evaluations \cite{Browne2013}.
As shown in Fig.~\ref{fig:chart}~b) this value for $^{255}$Fm deviates from all other known odd-mass \mbox{(odd-A)} nuclide and in fact, it is below the Schmidt line \cite{Schmidt1937, krane_introductory_1987} located at -1.913 $\mu_N$ and hence in unphysical space. This line is derived from a simplified single-particle shell model that assumes that the total nuclear magnetic moment equals the moment of its single unpaired nucleon \cite{deGroote2020_schmidt}.
To further elucidate this situation, we investigated the hyperfine splitting of atomic transitions in $^{255}$Fm with high resolution using the perpendicularly-illuminated laser ion source and trap (PI-LIST) \cite{PILIST2016} at the RISIKO mass separator \cite{Kieck2019}.
From the experimentally extracted hyperfine coupling constants, the nuclear magnetic dipole and electric quadrupole moments were determined by combining the experimental results with dedicated atomic and nuclear calculations, providing an updated and reliable reference for the nuclear moments.

The measurements were performed on several samples of $^{255}$Fm, each containing $10^8$ to $10^9$ atoms. 
These were produced from an initial sample of 8.8$\cdot$10$^{13}$ atoms of $^{254}$Es, equivalent to 34\,ng, provided by Oak Ridge National Laboratory (ORNL) and Florida State University, Tallahassee, USA \cite{oakridge2015}. This initial sample was neutron irradiated in the high-flux research reactor at the Institut Laue-Langevin (ILL) in Grenoble, France for 7\,days.  After a cooling period of 4\,days, the sample was transported to the Johannes Gutenberg University, Mainz, Germany. 7.5$\cdot$10$^{10}$ atoms of $^{255}$Es ($T_{1/2} = 39.8$\,d) \cite{NUBASE2020} were available to be used as a $^{255}$Es/$^{255}$Fm generator. This production path is depicted in red in Fig.~\ref{fig:chart}.
At the Department of Chemistry – TRIGA site at Mainz, on-site chemical separations were performed using a cation-exchange column chromatography technique with alpha-hydroxyisobutyrate as the eluting agent \cite{Warbinek2024}. The fermium fraction was then placed on a zirconium foil and evaporated to dryness. This on-site separation allowed performing an extended measuring campaign by yielding multiple samples at timed intervals after secular equilibrium was established between individual separations.

\begin{figure*}[tbp!]

    \includegraphics[width=1\textwidth]{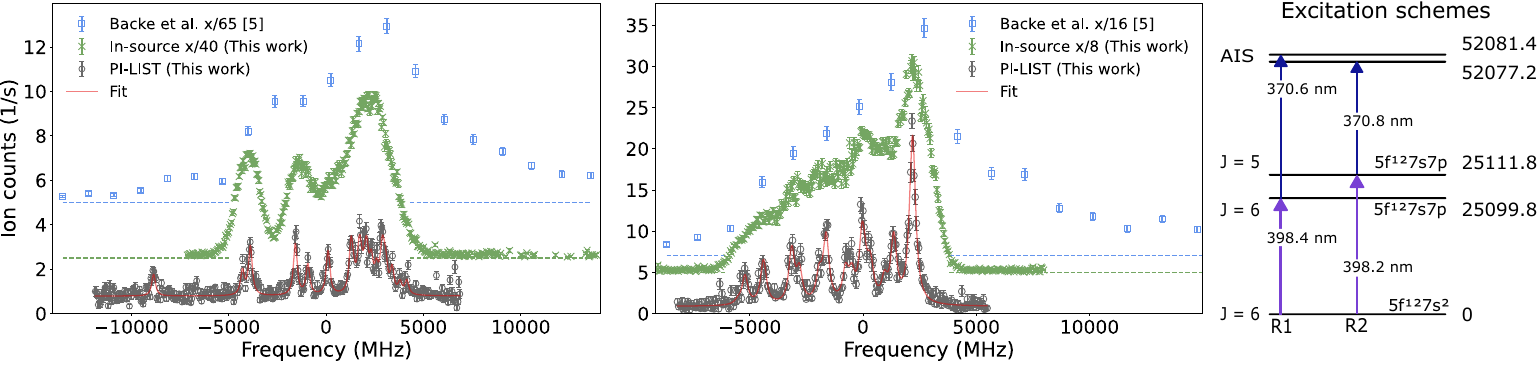} 
    
\caption{\label{fig:R1_plot} Measured HFS spectra for in-source spectroscopy (green) and with the PI-LIST (gray) together with results from Backe et al.~\cite{Backe2005}. For greater clarity, the spectra have an individual vertical offset (dashed lines). Left: spectra for R1 relative to the  energy of 25\,099.760(14)\,cm$^{-1}$. Center: spectra for R2 centered relative to the centroid energy of 25\,111.760(12)\,cm$^{-1}$. 
On the right, the excitation schemes and the auto-ionizing states (AIS) used in this work are shown, along with their energies expressed in wavenumbers (cm$^{-1}$). The atomic configurations of the GS and excited states are also indicated \cite{Allehabi2020}.}
\end{figure*}

At the RISIKO mass separator \cite{Kieck2019}, neutral fermium atoms were produced by placing a sample in the atomizer and resistively heating it to a typical temperature of 900$^{\circ}$C \cite{Kieck2019}. 
All laser excitations took place within the PI-LIST structure \cite{PILIST2016, PILIST_Kron_2020}, which consists of two repelling electrodes, installed immediately in front of the atomizer, and a radio frequency quadrupole (RFQ) ion guide. 
Within the RFQ structure, just a few millimeters apart from the front of the atomizer, the effused neutrals are intersected by the first laser beam. The spectroscopy (first step) laser is arranged in a perpendicular geometry and the ionization (second step) laser in an anti-collinear configuration relative to the effusing atomic beam.
The overlap of both lasers with each other and the atomic beam defines and limits the effective aperture angle of the atomic beam. This enables a narrow velocity spread with respect to the spectroscopy laser, reducing the Doppler broadening to approximately 50 MHz in the first step.

Ions created inside the RFQ structure were extracted through the PI-LIST's exit electrodes and accelerated to 30\,keV. 
They were then mass separated in a 60$^{\circ}$-sector-field dipole magnet according to their mass-to-charge ratio with a typical resolving power of M/$\Delta$M$\simeq$~600. 
Finally, ions were counted on a secondary electron multiplier (SEM) for single ion detection. 
By using pulsed lasers, a bunch structure was imprinted on the detected ions and a time-resolved acquisition further reduced the background, using a multichannel analyzer and gating to the bunch structure of 20-30\,\textmu s duration. 

Two pulsed titanium:sapphire (Ti:sa) laser systems provided the photons to drive two resonant transitions from the ground-state (GS), historically \cite{Sewtz2003, Backe2005} named \textit{Resonance 1} (R1) at $25\,099$\,cm$^{-1}$ and \textit{Resonance 2} (R2) at $25\,111$\,cm$^{-1}$, to the corresponding intermediate levels at the respective wavelengths as given in Fig.~\ref{fig:R1_plot}.
The narrow-bandwidth laser light of the first step was generated with an injection-locked Ti:sa laser \cite{Sonnenschein2017} and single-pass second harmonic generation (SHG), with typical average laser powers of 20\,mW in the PI-LIST interaction area. 
The typical spectral characteristics of this system were close to Fourier-limited bandwidths of 20\,MHz in the fundamental \cite{Sonnenschein2017} while the scanning range was defined by a continuous wave (cw) master laser. 
To ease operation, a diode laser (TOPTICA, DL PRO) was used as master laser for R1,  while for R2 a diode pumped cw-Ti:sa \cite{SONNENSCHEIN2020} was chosen. 
The ionizing pulses were generated by an in-house-built high-power Ti:sa laser with two gain crystals, providing 1.3\,W of laser light entering the vacuum chamber after SHG. 
The spectral, spatial, and temporal beam qualities were similar to those of a standard Ti:sa laser \cite{mattolat2009}. 
Both systems were driven by individual pump lasers at 532\,nm with a repetition rate of 10\,kHz, synchronized by an external pulse-pattern generator. 
The fundamental wavelength of both laser systems was constantly measured by a wavemeter (HighFinesse WSU30) with a nominal absolute accuracy of 30\,MHz, which was periodically calibrated to a well-known rubidium transition \cite{Verlinde}. Similar setups have been previously used for high-resolution spectroscopy of several elements \cite{Kaja2024, Weber2023, Nothhelfer2022, PILIST_Kron_2020, Studer2020}.

Recorded spectra in $^{255}$Fm for the transitions R1 and R2  are shown in Fig.~\ref{fig:R1_plot}. 
To ensure high efficiency, an initial signal was generated with the first-step laser in anti-collinear geometry, directly ionizing in-source. In this way, the first spectra were taken with limited spectral resolution due to the Doppler broadening, shown in green. The narrow-bandwidth first-step laser was then aligned perpendicularly in the PI-LIST, while being repeatedly scanned across the resonance structure, maintaining a substantially low but nearly constant atomic-beam current. 
These spectra, binned with 50\,MHz, are shown in gray, with the majority of the hyperfine components being well resolved and their best fit shown in red. Blue data points show data from Backe et al.~\cite{Backe2005} obtained from spectroscopy in a buffer gas cell. Here the spectral resolution was mostly limited by a strong Doppler broadening, pressure broadening, and a large laser bandwidth.

As the angular momentum $J$ of the atomic shell is larger than the nuclear spin $I = 7/2$, each atomic state splits into $2I + 1=8$ HFS levels.
With the achieved experimental resolution and efficiency, 15 out of 22 expected components were observed for R1 and 13 out of 21 for R2, while the remaining components were not fully resolved.
Nevertheless, the large redundancy in the structure allows extracting the hyperfine parameters with high precession.
A HFS model was fitted to the binned data using Satlas2 \cite{Satlas2}, knowing that the HFS energy splitting of an atomic level is given by
\begin{equation}\label{splitting}
    \Delta \nu = \mathcal{A} \cdot \frac{C}{2} + \mathcal{B} \cdot \frac{3 C (C + 1) - 4 I(I + 1) J (J + 1)}{8I(2I-1)J(2J - 1)},
\end{equation}
where $\mathcal{A}$ and $\mathcal{B}$ are the HFS coupling constants, and the Casimir factor $C = F(F + 1) - I(I + 1) - J(J + 1)$, where $F$ is the total angular momentum.
Derived from the electronic angular momentum ($J$) and nuclear spin ($I$), Racah intensities were calculated to determine relative strengths of the HFS components. 
Accordingly, fixed intensity ratios to close-lying poorly resolved peaks were assigned while the intensities of the fully-resolved components were set as free parameters to account for optical hyperfine pumping contributions and variations of the atom source conditions. 
Since the GS HFS coupling constants are the same for both transitions, these were shared between the two fitting models for a more precise result \cite{satlas}. 

The angular momentum $J$ of each respective excited state was also investigated, as a definite value had not been given before \cite{Backe2005}. As other assignments did not result in any convergence and did not lead to a reasonable reproduction of either HFS spectra, this work confirms the values for the exited levels $J_{\textrm{R1}_\textrm{e}} = 6$ and $J_{\textrm{R2}_\textrm{e}} = 5$ and thus, confirms the assignment to the levels given by Allehabi et al.~\cite{Allehabi2020}. 
A linewidth of 230 MHz (350 MHz) was extracted for R1 (R2), mostly dominated by power broadening, resulting from the compromise between resolution and efficiency. 
Taken from the best fit to the data as shown in Fig.~\ref{fig:R1_plot}, the extracted HFS coupling constants are summarized in Table~\ref{tab:AB_exp} where the GS values result from the shared fit model. 
Our results clearly differ from the values given in Ref.~\cite{Backe2005}, which were negatively affected by the low resolution and the presence of an unreproducible high-energy tail in their spectra, likely caused by their dye lasers' spectral profile. 
The high resolution of the PI-LIST spectra and the application in different geometries removes any ambiguity and revises these values with a significantly improved precision. 

\begin{table}
\caption{\label{tab:AB_exp}%
Experimentally extracted HFS coupling constants for $^{255}$Fm for each studied atomic energy level. R1$_\textrm{e}$ and R2$_\textrm{e}$ represent the first excited level for R1 and R2, respectively. An error of 50\% was estimated by Backe et al.~\cite{Backe2005}.}
\begin{ruledtabular}
\begin{tabular}{lcccccc}
\textrm{State}&
\textrm{$J$}&
\multicolumn{2}{c}{\textrm{$\mathcal{A}$ [GHz]}}&
\multicolumn{2}{c}{\textrm{$\mathcal{B}$ [GHz]}}\\

\textrm{}&
\textrm{}&
\textrm{This work}&
\textrm{Backe et al.}&
\textrm{This work}&
\textrm{Backe et al.}\\

\colrule
\textrm{GS}&
\textrm{6}&
\textrm{-0.1495(11)}&
\textrm{-0.32 }&
\textrm{-10.46(4)}&
\textrm{-22}\\

\textrm{R1$_\textrm{e}$}&
\textrm{6}&
\textrm{-0.3092(12)}&
\textrm{-0.53}&
\textrm{-12.79(4)}&
\textrm{-2.9}\\

\textrm{R2$_\textrm{e}$}&
\textrm{5}&
\textrm{-0.0097(13)}&
\textrm{-0.69}&
\textrm{-13.57(5)}&
\textrm{-1.7}\\

\end{tabular}
\end{ruledtabular}
\end{table}

Uncertainties were determined 
from statistical distribution of the ion counts and the systematic uncertainty inherent to the wavemeter, determined by modulating the experimental data with a triangle function with arbitrary phase as suggested in Ref.~\cite{Verlinde}. 
In addition, the dependence of the extracted results to the binning was investigated and found to be well within the extracted uncertainty range. 
As a cross check, the Doppler-broadened data were included in the fitting routine, resulting in a full agreement without significant influence on the final HFS values and their uncertainties.

To extract the magnetic dipole moment $\mu$ and the spectroscopic electric quadrupole moment $Q_\textrm{S}$, the knowledge of the hyperfine fields is needed. 
In the case of fermium, with no stable reference isotope aiding to fix atomic quantities with standard techniques, the magnetic field at the nucleus ($B_e(0)$) and the electric-field gradient of the electronic cloud at the site of the nucleus ($\langle \partial^2 V/\partial z^2 \rangle_{z=0}$) have to be taken from accurate atomic theory calculations \cite{Yang2023,Block2021}.
On the basis of the experimentally extracted values for $\mathcal{A}$ and $\mathcal{B}$ using equation \ref{splitting}, we obtain the nuclear moments for $^{255}$Fm from the relation 

\begin{equation}\label{eq:moments}
\mathcal{A} = \mu_I \frac{B_e(0)}{I J} \textrm{~and~ }
\mathcal{B} = eQ_\textrm{S}  \Biggl	\langle \frac{\partial^2 V}{\partial z^2} \Biggl \rangle_{z=0} ~~~ .
\end{equation}

The hyperfine fields in equation \ref{eq:moments} can also be equated for HFS coupling constants measured in different isotopes, enabling the derivation of nuclear moments if they are known for at least one isotope.

Despite the open $f$-shell character of fermium, recent atomic calculations using the configuration interaction with perturbation theory (CIPT) method predicted some neutral atomic level scheme along with hyperfine fields \cite{Allehabi2020, Dzuba2023}. To evaluate the reproducibility of these fields across different theoretical approaches, dedicated MultiConfigurational Dirac-Hartree-Fock (MCDHF) calculations were performed within this work.
The use of the MCDHF theory as implemented in \textsc{grasp18} \cite{FroeseFischer_GRASP2018_2019} is described in Appendix\,A. Table \ref{tab:ground_state} summarizes these results, including available results from calculations using CIPT methods as reported in Refs. \cite{Allehabi2020} and \cite{Stephan_JENA}, a previous MCDHF calculation reported by Ref.~\cite{test1}, and the new MCDHF results obtained in this work.
As the values show a very good agreement across the different calculations and different theoretical frameworks, the average value is taken for the extraction of nuclear parameters and its uncertainty calculated from the standard deviation of all four methods \cite{Barzakh2021, theo_errors2}. 

\begin{table}[t]
\caption{\label{tab:ground_state}%
Theoretical atomic calculations for the ground-state [Rn] $7\!s^2 5\!f^{12 } \ {}^{3}\!H_6$ hyperfine fields $\mathcal{A}/g_i$ and $\mathcal{B}/Q_\textrm{S}$ from CIPT and MCDHF theory. The values for the $\mathcal{A}$-factor are given in a nuclear spin-independent notation with respect to the $g_i$-factor with $g_i \cdot I=\mu_I$. Uncertainties are given only for the calculations done for this work. The average values are taken from all calculations and the uncertainties are their associated standard deviations \cite{Barzakh2021, theo_errors2}.} 
\begin{ruledtabular}
\begin{tabular}{lll}
\textrm{Method }&
$\mathcal{A}/g_i$ [GHz/$g_i$]&
$\mathcal{B}/Q_\textrm{S}$ [GHz/$e$b]\\
\colrule                 &           \\ [-2.3ex]
CIPT \cite{Allehabi2020} & 0.655      & -1.750      \\
CIPT \cite{Dzuba2023}    & 0.759      & -1.844     \\
MCDHF \cite{test1}       & 0.725      & -1.782      \\
MCDHF (This work)        & 0.694      & -1.789     \\ [0.5mm]  
\colrule                 &           \\ [-2.3ex]
Average                  & 0.708(44)  & -1.791(39) \\
\end{tabular}
\end{ruledtabular}
\end{table}
For the excited states R1 and R2 in the CIPT calculations \cite{Allehabi2020,Dzuba2023}, Dzuba et al.~\cite{Dzuba2023} declare that \textit{"the state R2 has [an] anomalously small value of $\mathcal{A}$. This means that the theoretical uncertainty for this state is large and the state is not very good for the extraction of nuclear parameters."} This is very much in agreement with our experimental findings and MCDHF results, indicating discrepancies and no convergence for either state. Thus, only the GS values are used further. 

The combination of the experimental results from Table\,\ref{tab:AB_exp} and the atomic calculations from Table\,\ref{tab:ground_state} enabled the extraction of nuclear moments for $^{255}$Fm as given in Table\,\ref{tab:nucl_parameters}. Here the magnetic dipole moment $\mu_I(^{255}\mathrm{Fm})=-0.743(49)\,\mu_{\mathrm{N}}$ is well within the Schmidt limit \cite{Schmidt1937}, in contrast to the evaluation performed with the unresolved spectra in Ref.~\cite{Backe2005}. 

To gain more insight, these results are compared with predictions obtained from two microscopic nuclear models based on the energy density functional framework. 
The first model is using the Gogny force D1M \cite{Goriely.2009} and consists in Hartree-Fock-Bogoliubov (HFB) calculations supplemented by approximate beyond-mean-field corrections \cite{Peru2021,Barzakh2021}.
The second model is using the Skyrme SLyMR1 interaction \cite{Sadoudi2013,JodonPHD} and performs a full configuration mixing of symmetry-restored triaxial reference states within a Multi-Reference Energy Density Functional (MREDF) approach \cite{Bally2014,Bally2022}. Details about the models can be found in Appendix\,B. We remark that both models also describe the evolution of isotopic shifts of charge radii of fermium isotopes reasonably well \cite{Warbinek2024}.

\begin{table}[t]
\caption{\label{tab:nucl_parameters}%
Extracted experimental nuclear parameters for $^{255}$Fm in comparison to nuclear calculations (bold). The theoretical calculations cover further odd fermium isotopes and isomers, as indicated. Tentative ground state spin assignments (indicated by parentheses) are used to infer the level from the nuclear calculations.} 
\begin{ruledtabular}
\begin{tabular}{rclll}
Isotope & $I/\hbar$ & \textrm{Method }  & $\mu / \mu_\textrm{N}$  &  $Q_\textrm{S}$/$e$b \\
\colrule & \\[-2.2ex]
\textbf{$^{255}$Fm}  & 7/2   & Experiment & \textbf{-0.743(49)} & \textbf{5.84(13)} \\
\\[-1.5ex]
$^{257}$Fm           & (9/2) & MREDF     & \,\,\,1.7                 & 7.7               \\
\textbf{$^{255}$Fm}  & 7/2   & MREDF     & \textbf{\,-1.0}         & \textbf{6.2}      \\
\\[-1.5ex]
$^{257}$Fm           & (9/2) & HFB       & \,\,\,1.4                 & 6.67              \\
\textbf{$^{255}$Fm}  & 7/2   & HFB       & \textbf{-0.78}      & \textbf{5.78}     \\
$^{253}$Fm           & 1/2   & HFB       & \,-1.047              & 0                 \\
$^{251}$Fm           & (9/2) & HFB       & \,-0.730              & 7.06              \\
$^{249}$Fm           & (7/2) & HFB       & \,\,\,1.215               & 6.01              \\
$^{247\textrm{g}}$Fm & (7/2) & HFB       & \,\,\,1.211               & 5.99              \\
$^{247\textrm{m}}$Fm & (1/2) & HFB       & \,\,\,0.363               & 0                 \\
$^{245}$Fm           & (1/2) & HFB       & \,\,\,0.355               & 0                 \\

\end{tabular}
\end{ruledtabular}
\end{table}

These nuclear calculations agree very well with the experimental results as summarized in Table\,\ref{tab:nucl_parameters}. 
The MREDF calculations overestimate the nuclear magnetic dipole moment by approximately 25\% and the electric quadrupole moment by around 6\%, whereas the deviation from the HFB results remains below 5\% in both cases.
In addition, the nuclear moments of $^{255}$Fm closely match the values of the isotone even-odd nucleus $^{253}$Cf with $\mu_I(^{253}\mathrm{Cf})=-0.731(35)\,\mu_{\mathrm{N}}$ and $Q_\mathrm{S}(^{253}\mathrm{Cf})=5.53(51)$\,$e$b \cite{Weber2023}. This indicates that the same neutron valence orbital is occupied in both nuclei, suggested to be the $\nu$\,7/2[613] Nilsson orbital for $^{253}$Cf \cite{Browne2013}.
This orbital is then also expected to be involved in the odd-odd nucleus $^{254}$Es, where it couples with the $\pi$\,7/2[633] orbital (valence proton in $^{253,255}$Es) in a full alignment to a total spin of $I_{^{254}\mathrm{Es}}=7$ \cite{Nothhelfer2022}. Within their uncertainty, the quadrupole moments are also identical between $^{255}$Fm and $^{253}$Cf, which is expected in this region of stable, strongly prolate deformed nuclei where experimental charge radii did not reveal any change in shape \cite{Warbinek2024}.

\begin{figure}
\includegraphics[width=0.45\textwidth]{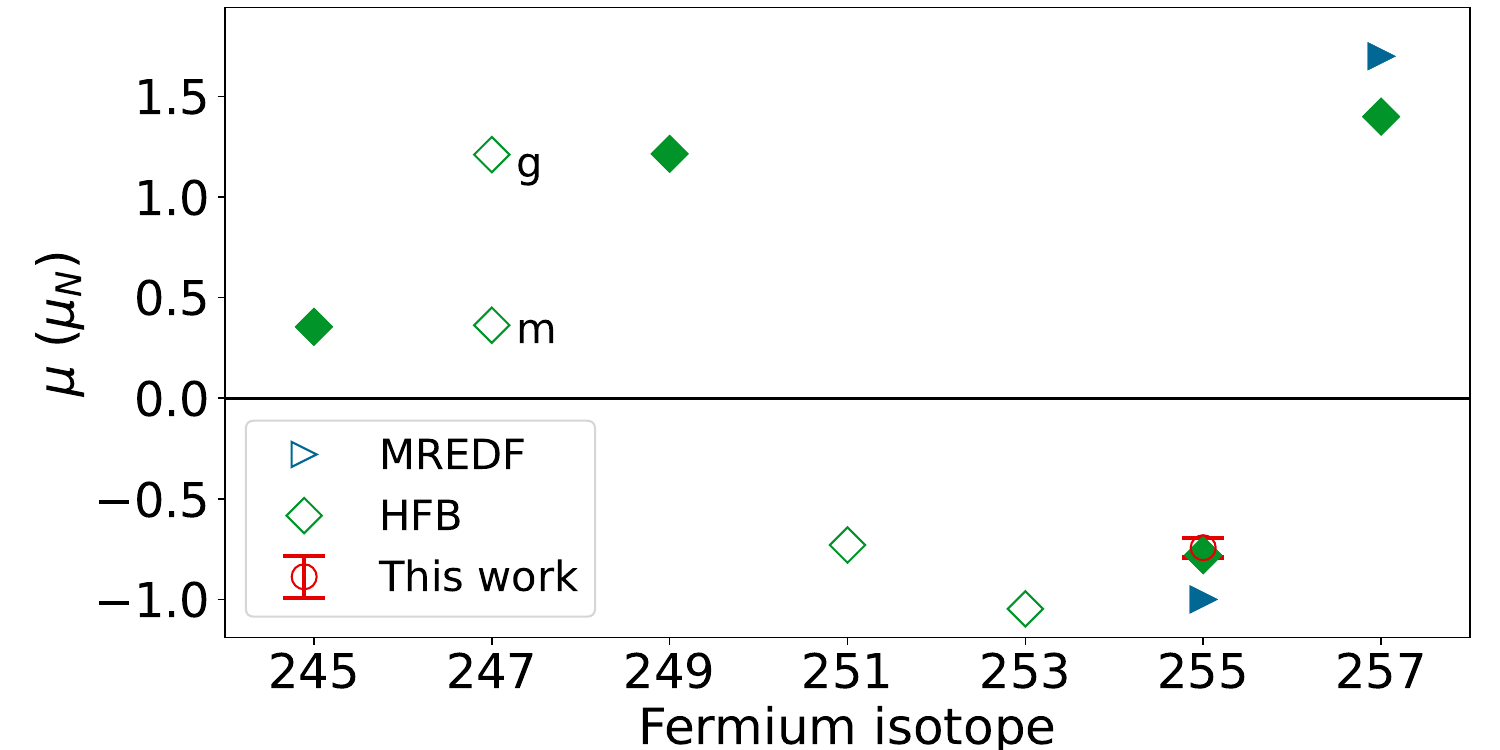}
\caption{\label{fig:moments} Predicted and experimental magnetic moments for some fermium isotopes. Common isotopes also included in Ref.~\cite{Warbinek2024} are marked as solid symbols.}
\end{figure}

Based on these nuclear structure calculations, in Fig.~\ref{fig:moments}, we further provide predictions for the magnetic dipole moment of the ground state of neighboring fermium isotopes where experimental results from high-resolution laser spectroscopy are not yet available. 
However, several isotopes were studied with reduced resolution recently by Warbinek et al.~\cite{Warbinek2024}. As the total width of the HFS is determined by the magnetic dipole moment, a qualitative comparison can be made with these results. From their appendix section, 
the HFS of $^{257}$Fm is clearly wider than that of $^{245}$Fm, which is in agreement with the prediction made by the HFB calculations.

In this Letter we have reported on the first high-resolution spectra of an atomic transition in the isotope $^{255}$Fm, using the PI-LIST at the RISIKO mass separator, with samples of around $10^8$ atoms. 
This enabled probing the HFS of atomic transitions, previously only observed unresolved \cite{Sewtz2003,Backe2005}. 
In combination with reported and newly calculated atomic coupling constants from atomic theory the nuclear magnetic dipole moment $ \mu_I (^{255}\mathrm{Fm})  = -0.75(5)\,\mu_\textrm{N}$ and the nuclear electric quadrupole moment $ Q_S (^{255}\mathrm{Fm}) = +5.84(13)$\,$e$b were determined. 
These results were compared to predictions from nuclear MREDF and HFB calculations, which are in good agreement for $^{255}$Fm. The nuclear theory calculations covered additional odd-A fermium isotopes, some of which have previously been studied in on-line experiments with limited resolution \cite{Warbinek2024}. Improved experimental results appear within reach, e.g., from future upgrades of the JetRIS apparatus at GSI Darmstadt \cite{JetRIS} and the upcoming S$^3$ low-energy-branch at GANIL, France \cite{romans2022}. $^{255}$Fm now serves for the prediction of expected HFS and is available as a reference isotope.

\begin{acknowledgments}
\textit{Acknowledgments} \textemdash \,
We thank the mechanical workshop and radiation protection staff at TRIGA Mainz and R. Jera, glassblower at JGU Chemistry, the ILL reactor and health physics teams, and N. Sims, radioisotope laboratory technician at ORNL. The isotopes used in this research were supplied by the U.S. Department of Energy, Office of Science, Isotope R\&D and Production program. This research was supported by the U.S. Department of Energy Office of Science, Office of Basic Energy Sciences, Heavy Element Chemistry program under Award Number DE-SC0023693. The $^{253,254,255}$Es was provided to Florida State University via the Isotope Development and Production for Research and Applications Program through the Radiochemical Engineering and Development Center at ORNL. S.G. acknowledges financial support from F.R.S.-FNRS (Belgium). This work has received funding from the German Bundesministerium für Bildung und Forschung under Project No. 05P21UMFN3, 05P15RDFN1 and 05P21RDFN1; from the Swedish Research Council under Grant No. 2024-04152; from the European Research Council under the European Union’s Horizon 2020 Research and Innovation Programme (Grant Agreement No.\ 819957) and Horizon Europe Research and Innovation Programme (Grant Agreement No.\ 101162059). This Marie Sklodowska-Curie Action Innovative Training Networks received funding from the European Union’s H2020 Framework Programme under grant agreement No. 861198. 

\end{acknowledgments}


\bibliography{apssamp}

\clearpage
\appendix
\textit{Appendix A: Atomic theoretical methods}\label{AppendixA} \textemdash \, 
Calculations involving open $f$-shell systems are computationally very
demanding, especially for excited states, therefore the focus in this investigations was on the ground state 
[Rn] $7s^2 5f^{12} \ {}^{3}$H$_6$ GS.
In particular, calculations involving MCDHF theory as implemented in \textsc{grasp18} \cite{FroeseFischer_GRASP2018_2019} were performed for this work.
\ja{%
    In the standard formulation of the MCDHF theory, the function $\Psi $ for a given state is represented as a linear combination of symmetry-adapted configuration state functions (CSFs),%
}
\begin{equation*}
    \Psi \left( \Gamma \pi J M \right) = \sum_{i=1}^{N_{\text{CSF}}} c_{i} \Phi \left( \gamma_{i} \pi J M \right),
\end{equation*}
where $J$ is the total electronic angular momentum, and $M$ is its projection, $\pi$ is the parity, $\gamma_{i}$ represents the set of orbital occupancies and complete coupling tree of angular quantum numbers unambiguously specifying the $i$-th CSF, while $\Gamma$ denotes all other information needed to uniquely describe the state \cite{Jonsson_Introduction_2022}.
A single reference calculation was performed on the ground state of Fm\,I. Five layers of correlation orbitals were generated by allowing single and double (SD) substitutions from the 5$f$ and 7$s$ valence orbitals. The maximum orbital angular momentum quantum number of the correlation orbitals was $\ell \le  5 $, which includes the $g$ correlation orbitals to properly account for correlation effects between electrons in $f$ shells. 
\ja{Afterwards, the relativistic configuration interaction \textsc{rci} program was run to increase the active set of orbitals while keeping the orbital shapes fixed. In this stage, configuration state functions (CSFs) were generated by substituting orbitals with a principal quantum number $n \ge 4$. For this, a restriction of single and restricted double (SrD) substitutions was applied which means at most two electrons may be substituted with at minimum one belonging to the valence orbitals (5$f$, 7$s$).}
In this stage, configuration state functions (CSFs) were generated by allowing substitutions of orbitals with principal quantum numbers $n \ge 4$. The substitutions were restricted to single and restricted double (SrD) excitations, meaning that at most two electrons could be substituted, with at least one electron coming from the valence orbitals (5$f$, 7$s$).

\ja{The calculations were repeated for the lanthanide homologue Er\,I to demonstrate the predictive accuracy of our model. Good agreement was found between previous theory and experiment for Fm\,I and Er\,I, respectively.}
\ja{Convergence was demonstrated for the HFS parameter $\mathcal{A}$ and  at least weak convergence demonstrated for the HFS parameter $\mathcal{B}$ in both Er\,I and Fm\,I. Nevertheless, further work might still be needed to demonstrate solid convergence for the $\mathcal{B}$ HFS parameter.}
\ja{This stage of the calculation represents one of the largest MCDHF calculations of a single state. It required generating 16 million CSFs, reduced to 9 million after removing those CSFs which did not directly interact with the reference list~\cite{Jonsson2023}}.

\textit{Appendix B: Nuclear theoretical methods}\label{AppendixB} \textemdash \,
%
%
The two microscopic nuclear structure models used to interpret the experimental findings
are based on an energy density functional (EDF) framework.  
They both describe reasonably well the evolution 
of the mean squared charge radii 
of fermium isotopes \cite{Warbinek2024}. 
One is using the Skyrme interaction SLyMR1 \cite{Sadoudi2013,JodonPHD}, 
and the other the Gogny force D1M  \cite{Goriely.2009}. 
The latter model has also already been used to describe nuclear moments of 
californium isotopes \cite{Weber2023}.
In either case, the starting point is the calculation of energy surfaces 
of several self-consistently blocked one-quasiparticle (qp) configurations 
variationally optimized through the Hartree-Fock-Bogoliubov (HFB) method, 
followed by a selection of the configurations that are compatible with the 
experimentally deduced spin and parity. 
Contrary to the spectroscopic quadrupole moment, the magnetic moment 
of a deformed nucleus cannot be directly calculated as the expectation 
value of the magnetic dipole  operator of a qp configuration. Instead, 
the axial HFB calculations with D1M are mapped on the unified model 
\cite{BMII} as described in Ref.~\cite{Peru2021}. In addition, to approximately 
take configuration mixing into account, the D1M results average over 
deformation, as described in Ref.~\cite{Barzakh2021}.
By contrast, for SLyMR1 calculations, a full
configuration mixing of symmetry-restored reference states is performed
within a Multi-Reference Energy Density Functional (MREDF) framework including
triaxial shapes as described in Ref.~\cite{Bally2022}.
Finally, we mention that the SLyMR1 results for magnetic moments are calculated with bare $g$ factors, 
whereas for the D1M results the spin contributions are quenched 
with a factor of 0.75 \cite{Peru2021}.  

\end{document}